\documentclass[manuscript]{aastex}

\usepackage{graphicx,natbib}
\usepackage[percent]{overpic}

\def\Rs{R_{\odot}}

\slugcomment{}

\shorttitle{Homologous CMEs}
\shortauthors{Chatterjee \& Fan}

\begin{document}

\title{Simulation of homologous and cannibalistic
Coronal Mass Ejections produced by the emergence of a twisted flux rope into the Solar Corona}

\author{Piyali Chatterjee and Yuhong Fan}
\affil{High Altitude Observatory, National Center for Atmospheric Research, 3080 Center Green Drive, Boulder, CO 80301, USA}

\email{mppiyali@ucar.edu}

\begin{abstract}
We report the first results of a magnetohydrodynamic (MHD) simulation of the development of
a homologous sequence of three coronal mass ejections (CMEs) and demonstrate their so-called cannibalistic behavior.
These CMEs originate from the repeated formations and partial eruptions of kink unstable flux ropes as a 
result of continued emergence of a twisted flux rope across the lower boundary into a pre-existing coronal potential arcade field.
The simulation shows that a CME erupting into
the open magnetic field created by a preceding CME has a higher speed.  The second of the three
successive CMEs is cannibalistic, catching up and merging with the first into a single
fast CME before exiting the domain. All the CMEs including the leading merged CME, attained speeds
of about 1000 km s$^{-1}$ as they exit the domain. The reformation of a twisted flux rope after each CME
eruption during the sustained flux emergence can naturally explain the X-ray observations of
repeated reformations of sigmoids and ``sigmoid-under-cusp" configurations at a low-coronal source of homologous CMEs.
\end{abstract}


\section{Introduction}
Emerging solar active regions with strong photospheric magnetic twist
are known to repeatedly flare and produce homologous
coronal mass ejections \citep[e.g.][]{Gibson:etal:2002,Schrijver:2009}.
Some of these regions exhibit continued sunspot rotations
over the period of occurances of multiple flares
\citep[e.g.][]{Gibson:etal:2002,Brown:etal:2003,Zhang:etal:2008,Vemareddy:etal:2012},
which may indicate continued emergence of highly twisted flux
ropes from the interior into the corona \citep[e.g.][]{Schrijver:2009,Fan:2009,Fang:etal:2012}.
\cite{Gibson:etal:2002} studied AR 8668 throughout its passage on
the solar disk and found repeated reformation of the soft X-ray sigmoid
morphology after every filament eruption and temporarily parts 
of the sigmoid transform into cusp shape. 
Homologous CMEs have been observed to 
give rise to ``cannibalism'' or CME-CME interactions which are one of the
most energetic and geo-effective 
space weather phenomena \citep{Gopalswamy:etal:2001}. 
Such cannibalism events taking place in the solar wind in the heliosphere
have been modeled by \cite{Lugaz:etal:2005}, by launching identical CMEs
with the introduction of non equilibrium flux ropes in the lower corona at 
appropriate times. MHD simulations of homologous eruptions have been
achieved by \citet{Devore:Antiochos:2008} using a breakout magnetic
configuration in the corona driven by twisting footpoint motions,
although the resulting eruptions are confined. In this letter we present a 3D
MHD simulation of the initiation of homologous CMEs in the corona driven at
the lower boundary by the quasi-static emergence of a highly twisted
magnetic torus.  We observe cannibalism of the erupting CMEs as well as
repeated reformation of the helical flux rope after every eruption. 

\section{The Numerical Model}
In this simulation, we solve the MHD equations in
spherical geometry as given in \citet[][here after F12]{Fan:2012}, except that
here we exclude the field aligned thermal conduction term from the energy equation.
We have assumed an ideal gas with a low adiabatic index of
$\gamma = 1.1$, which allows the coronal plasma to maintain its high temperature
without an explicit coronal heating.
The MHD equations are solved numerically with the MFE code described in F12.
The setup of the simulation is the same as that of F12
except for the changes described below.
The spherical simulation domain is
given by $r \in [R_{\odot}, 6 R_{\odot}]$ with $R_{\odot}$ being the solar radius,
$\theta \in [ 11 \pi /24, 13 \pi / 24]$, $\phi \in [- \pi /12.8, \pi /12.8]$.
The domain is resolved by a grid of $480 \times 192 \times 360$, which is
uniform in $\theta$ and $\phi$, and non-uniform in $r$ with the highest resolution
being $dr = 0.95$ Mm in region from $r= R_{\odot}$ to $r=1.41 R_{\odot}$, and then
with $dr$ increasing gradually for $r>1.41 \, R_{\odot}$, reaching about
$dr=0.11 \, R_{\odot}$ at the outer boundary.
Compared to F12, here we have reduced the horizontal extent of the
simulation domain by about a factor of 2 and increased the resolution in the lower
coronal region by about the same factor, with the aim to model a more compact CME
source region with a stronger coronal magnetic field, representative of a highly twisted,
strong emerging active region that is capable of producing fast CMEs.
As in F12, the domain is set to be initially in hydrostatic equilibrium with a uniform
temperature of $T_0 = 1$ MK, and contains a pre-existing potential arcade field, whose
normal field distribution at the lower boundary is as given in F12 (see eqs. [12] and [13])
in that paper), except that the parameter for the width of the arcade normal field is
reduced to $\theta_a = 0.025$ and the peak field strength is increased to $B_0 = 70$ G.

As described in F12, we impose (kinematically) at the lower boundary
(at $r=R_{\odot}$) the emergence of a twisted magnetic torus
${\bf B}_{\rm tube}$, by specifying a time dependent transverse electric field
${\bf E}_{\perp}|_{r=R_{\odot}} = {\hat{\bf r}} \times \left [ \left (
- \frac{1}{c} \, {\bf v}_0 \times
{\bf B}_{\rm tube} \right )
\times {\hat{\bf r}} \right ]$ that corresponds to the upward advection
of the torus at a velocity ${\bf v}_0$.
The form of the torus ${\bf B}_{\rm tube}$ is given in F12 except that here it
is more compact, tightly wound, and has a stronger field, with
the minor and the major radii $a = 0.0212 R_{\odot}$ and $R' = 0.11 R_{\odot}$,
the twist rate $q/a = 0.142$ rad ${\rm Mm}^{-1}$, and the field strength at the
torus axis $B_t a/R' = 93 $ G. 
This field strength chosen is such that the emerging rope can be
confined by the arcade 
field during the quasi-static build up phase yet strong enough to produce an
ejective eruption (instead of a confined one) during the loss of equilibrium
phase and is also consistent with the typical
coronal field strength in strong active regions that produce large
flares \citep{Caspi:Lin:2010}.
Note that the outer poloidal field of the emerging flux rope has nearly the
same orientation as the arcade field in order to minimize magnetic reconnection
during the quasi-static build up phase.
The field of the torus ${\bf B}_{\rm tube}$ is truncated to zero outside of the
flux surface whose distance to the torus axis is $2a$.
For specifying the lower boundary electric field
${\bf E}_{\perp}|_{r=R_{\odot}}$, it is assumed that at $t=0$
the torus' center is located at a distance of $R'+2a$ below the lower boundary
(with the torus outer edge just approaching the lower boundary),
and it moves upward at a constant emergence speed of
$v_0 = 4.9$ km s$^{-1}$. This speed is much smaller than the peak Alfv\'en speed
$v_{A0} = 6.8 $ Mm s$^{-1}$ at the footpoints of the arcade field, and also
significantly smaller than the (initial) sound speed
$c_{s0}=135$ km s$^{-1}$ of the corona.
The emergence is thus sufficiently slow such that the fast coronal Alfv\'en
speed can quickly establish equilibria and allow the coronal magnetic field to
evolve quasi-statically in response to the driving flux emergence at the lower
boundary, until instabilities and catastrophic loss of equilibrium take place. 
Conversely, we may not reduce the emergence speed too much so as to avoid
a significant numerical dissipation of the flux rope current during
the quasi-static phase.
The other boundary conditions are the same as those used in F12, where we assume
perfectly conducting walls for the side boundaries, and use a simple outward
extrapolating boundary condition for the top boundary that allows plasma and
magnetic field to flow through.

\section{Results}
\begin{figure}[h!]
\begin{overpic}[width=0.22\textwidth,clip]{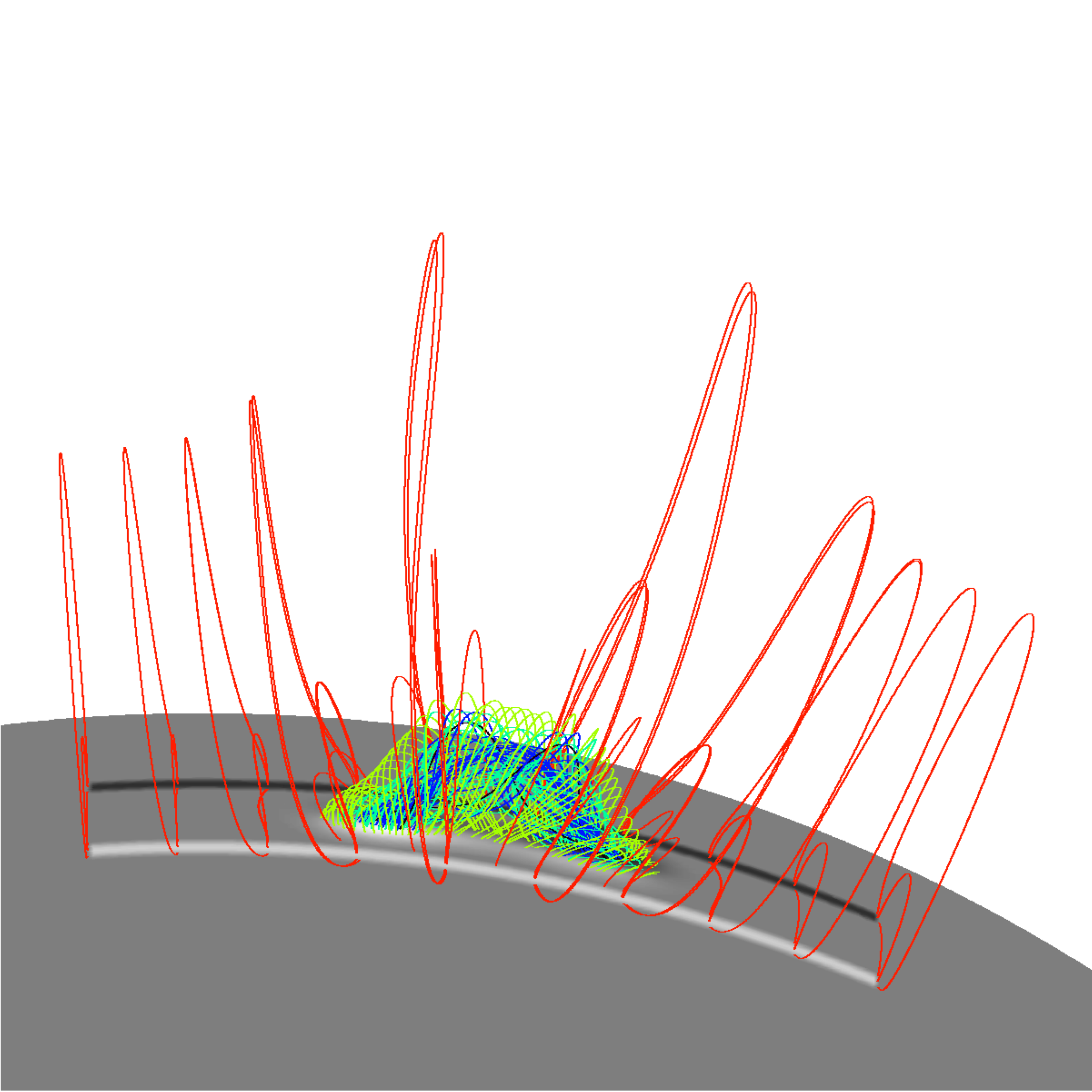}
\put(10,8){(a) $t=2.62$ }
\end{overpic}
\begin{overpic}[width=0.22\textwidth,clip]{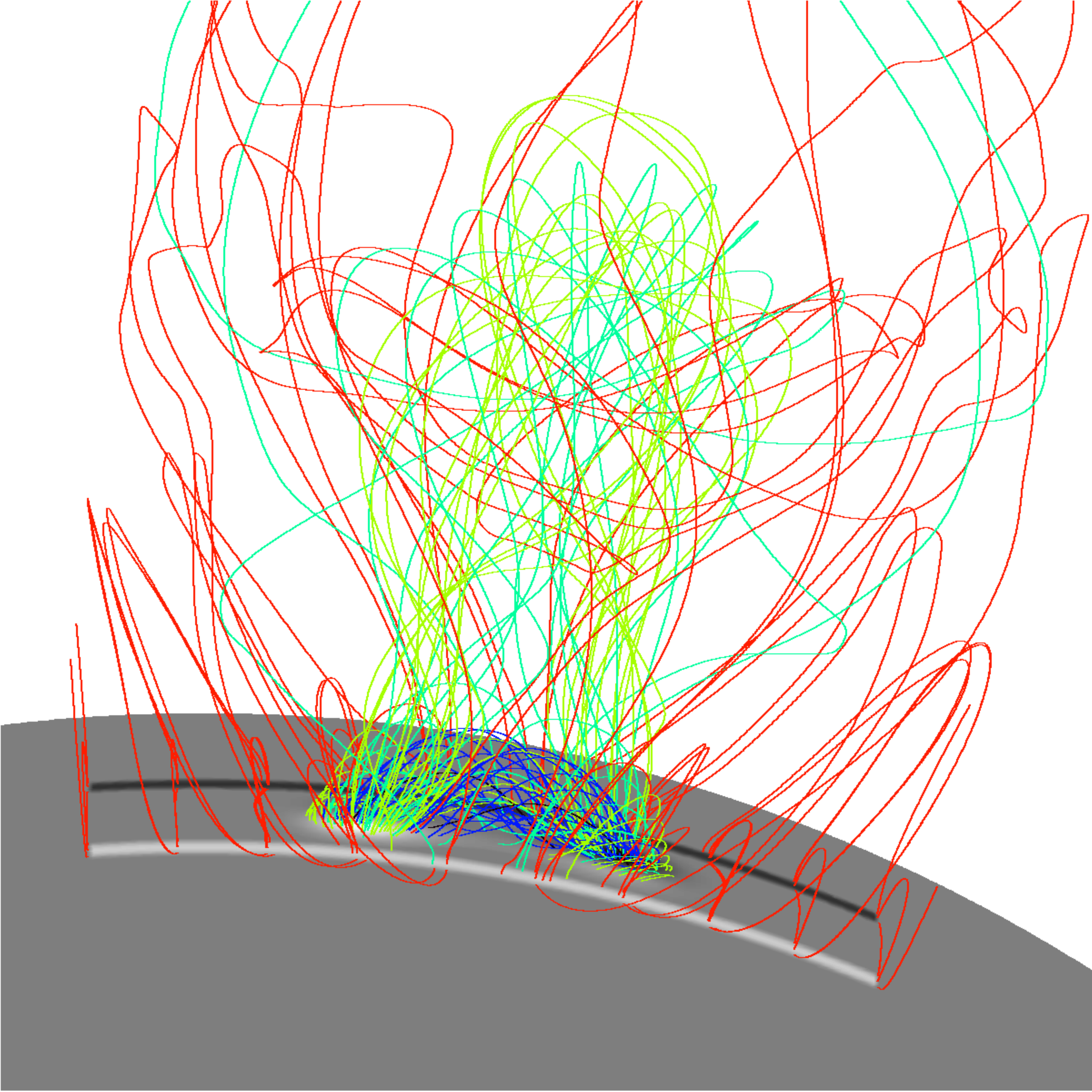}
\put(10,8){(e) $t=3.09$ }
\end{overpic}\\
\begin{overpic}[width=0.22\textwidth,clip]{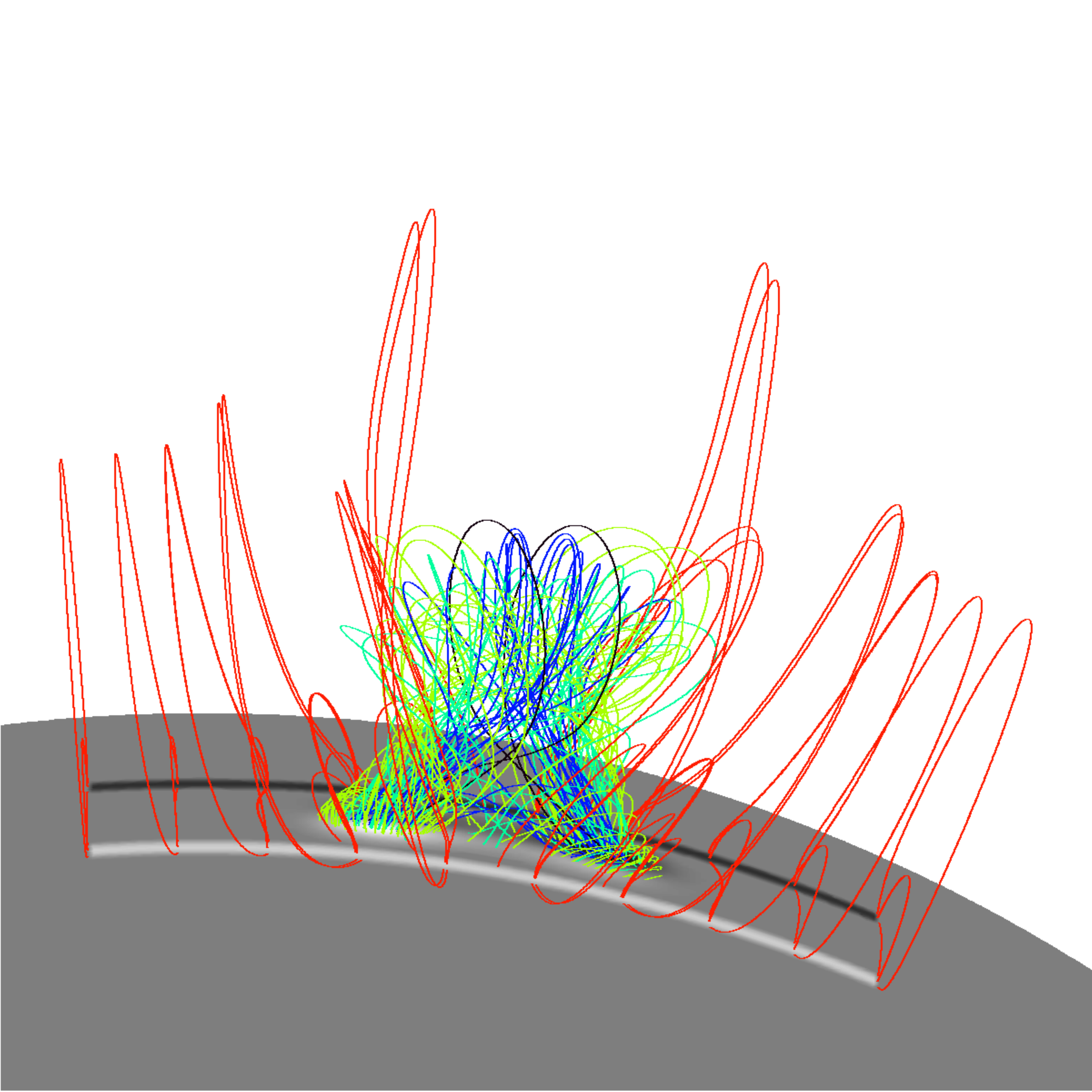}
\put(10,8){(b) $t=2.72$ }
\end{overpic}
\begin{overpic}[width=0.22\textwidth,clip]{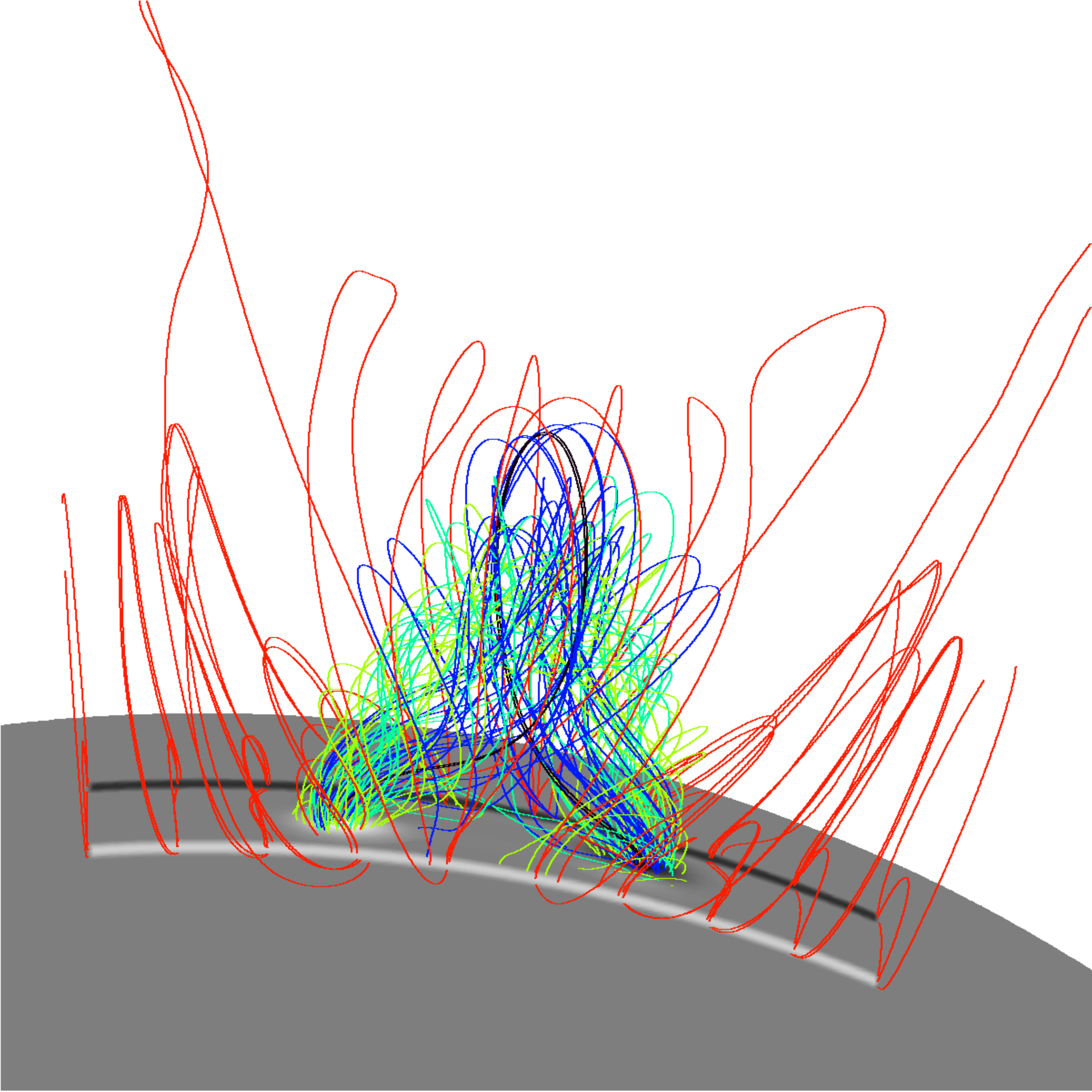}
\put(10,8){(f) $t=3.66$ }
\end{overpic}\\
\begin{overpic}[width=0.22\textwidth,clip]{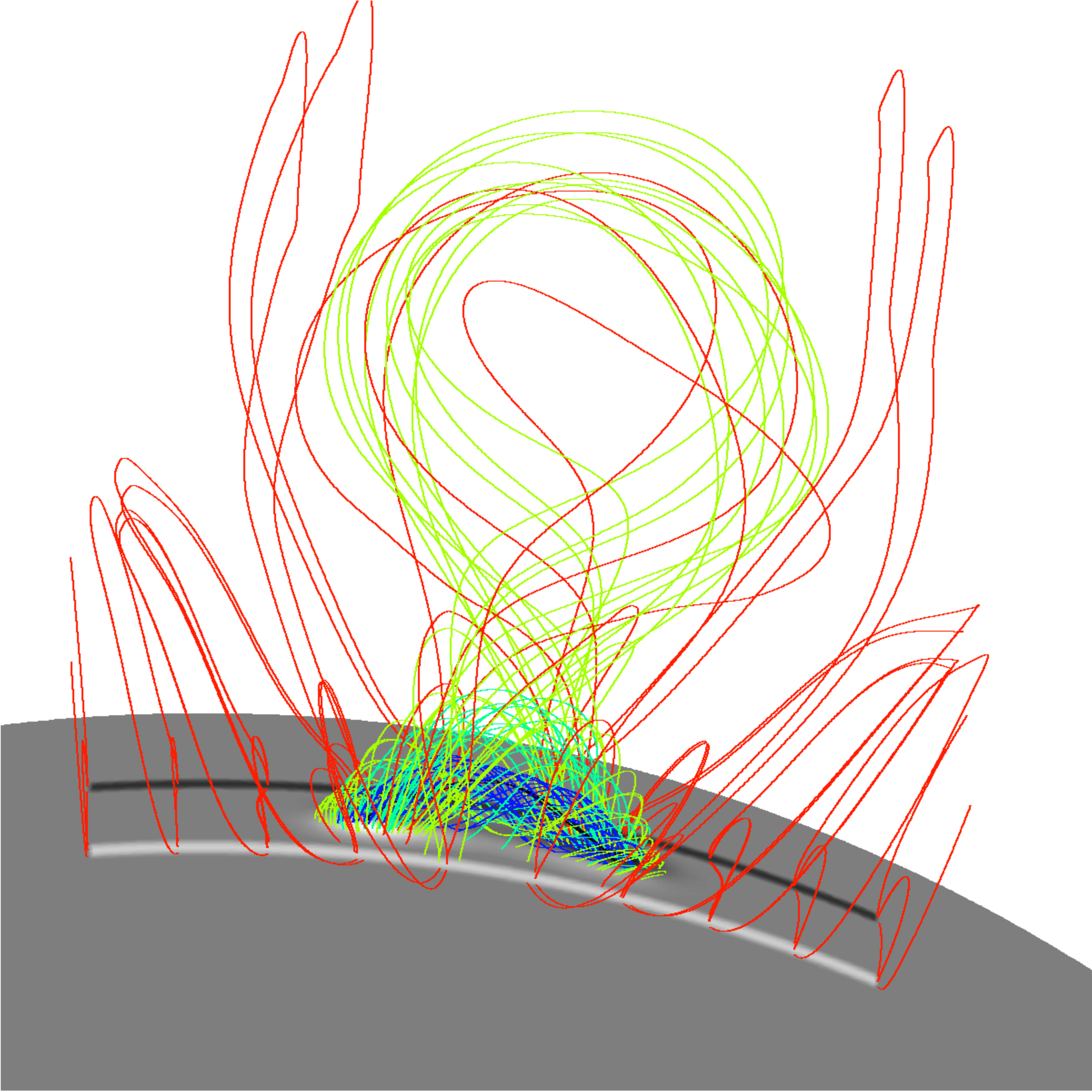}
\put(10,8){(c) $t=2.82$ }
\end{overpic}
\begin{overpic}[width=0.22\textwidth,clip]{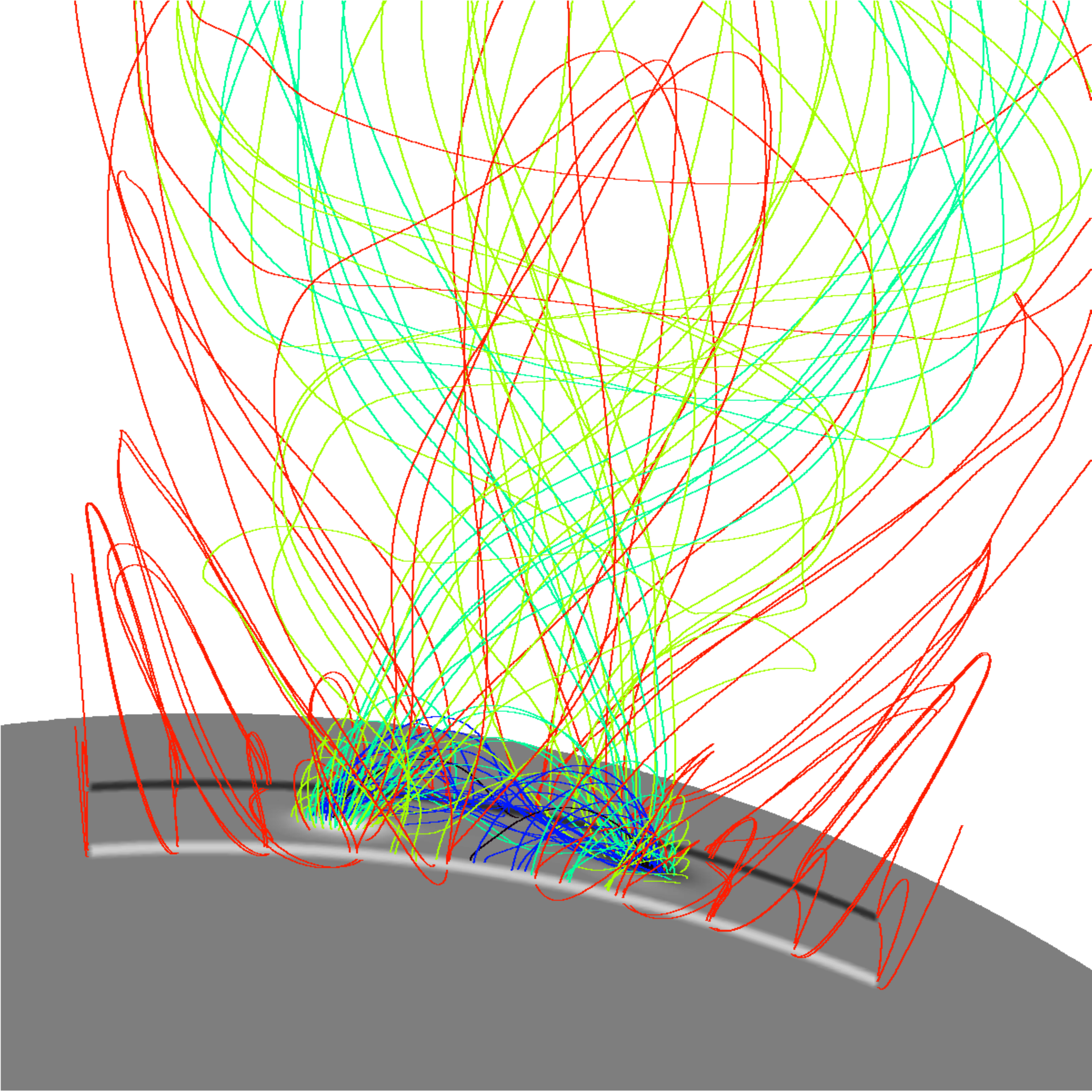}
\put(10,8){(g) $t=3.74$ }
\end{overpic}\\
\begin{overpic}[width=0.22\textwidth,clip]{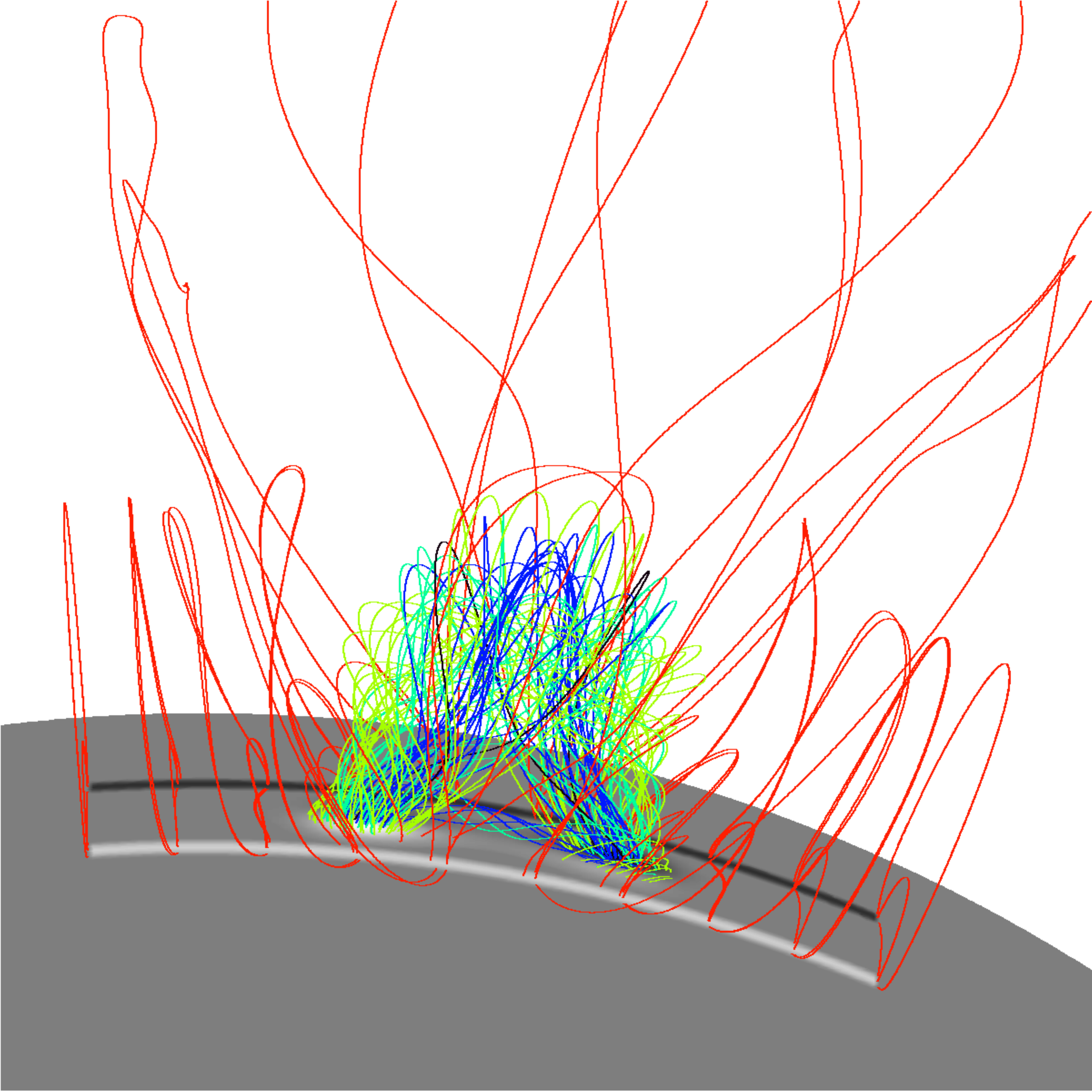}
\put(10,8){(d) $t=3.04$ }
\end{overpic}
\begin{overpic}[width=0.22\textwidth,clip]{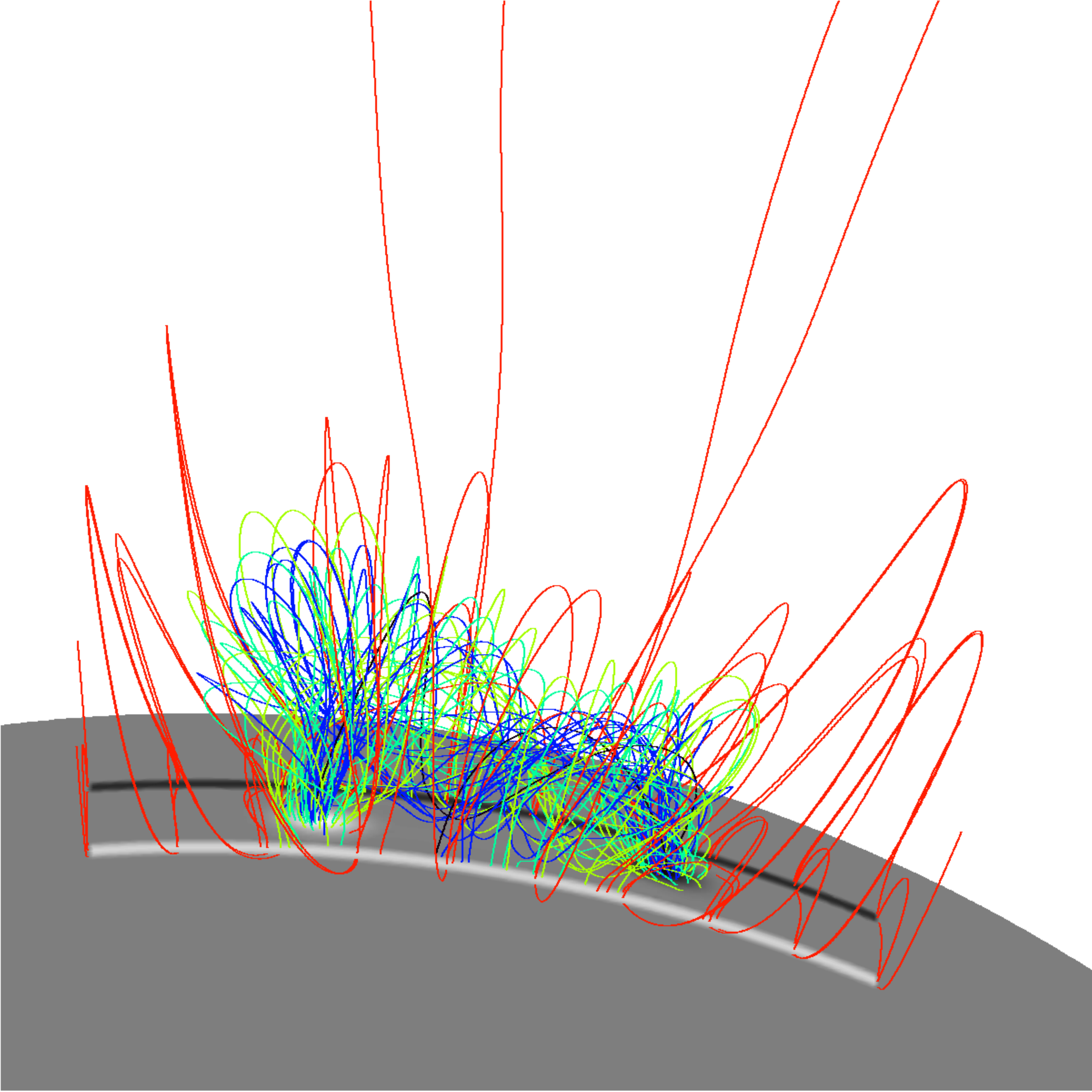}
\put(10,8){(h) $t=4.95$ }
\end{overpic}
\caption{\label{fig:snaps} The 3D magnetic field evolution of the twisted flux rope emerging into the corona at times indicated in hours. The red coloured field lines have foot points in the ambient arcade where as the blue, green and cyan field lines originate from the emerging flux region. A movie of the evolution is available in the online paper.}
\end{figure}

Our simulation shows repeated formation and eruption of a coronal flux rope
(at least 3 times) during the course of about 5 hours of evolution, as a result of
the continued emergence of the magnetic torus imposed at the lower boundary.
Fig.\ref{fig:snaps} shows snapshots of the 3D magnetic field evolution illustrating
this pattern of evolution in the corona. Panel (a) shows the emerged coronal flux rope
just before the onset of the first eruption.
The first eruption is found to initiate when
the twist of the emerged flux rope field lines in the 
vicinity of the axis has
reached about 2.1 winds between the anchored foot points.
The helical kink instability is expected to develop for a line-tied coronal
flux rope if the total winds of the field line twist about the axis exceed a critical
value between the line-tied ends \citep[e.g.][]{Hood:Priest:1981, Toeroek:Kliem:2003,
Toeroek:etal:2004}.
This critical value is 1.25 based on the analytical
calculation of a 1D uniformly twisted cylindrical flux tube \citep{Hood:Priest:1981}.
Fig.\ref{fig:snaps}b clearly shows the development of substantial writhing motion
at the onset of the eruption, indicative of the onset of the helical kink
instability.
Further, we check if the flux rope at this time is 
also unstable to the torus instability \citep{Bateman:1978, Kliem:Toeroek:2006,
Isenberg:Forbes:2007}, an expansion instability of a flux rope that occurs when the
external strapping field confining the flux rope decreases with height, $h$,
above the surface at a sufficiently steep rate.
The external strapping field is taken 
to be the potential field $B_{\rm P}$ with the same normal field distribution at the
lower boundary. 
The rate of decline with $h$ is measured by the decay index $n = -d\ln B_{\rm P}/d\ln h$.
We calculate this decay index at the apex of the axial field line of the
coronal flux rope at the start of the acceleration phase, which we estimate to be
at $r=1.035\Rs$. 
The apex of the flux rope axis is determined using the technique outlined
in \S3.1 of \cite{Fan:2010}.
The critical rate of decline for the onset of the torus instability
is determined to be $n_{\rm cr} = 1.5$ \citep{Bateman:1978} for a freely expanding 2D axisymmetric
toroidal current. The $n_{\rm cr}$ value for a 3D line-tied arched flux rope has been calculated
by \citet{Isenberg:Forbes:2007} and is found to be close to 1.5. 
In general, $n_{\rm cr}$ and the critical height are expected to depend on
the detailed normal
flux distribution at the lower 
boundary as well as the profile of the flux rope. \cite{Aulanier:etal:2010} found that their flux rope becomes unstable after reaching a critical height at which $n_{\rm cr} \sim 1.5$ where as \cite{Fan:2010} found a value of $1.74$ for $n_{\rm cr}$.
In our case the rope is probably stable against the torus instability at
the onset of the kink instability since we find a decay index of $n\sim 1.0$
at the apex of the rope axis when it has started to kink and accelerate
rapidly.
Subsequently in Fig.\ref{fig:snaps}c, we see the top of 
the erupting rope pinching off via magnetic reconnections and a new
second flux rope has formed due to
further flux emergence. The new flux rope constitutes not only sigmoid-shaped dipped field lines left over from the first eruption but also additional twist due to continued flux emergence. The second flux rope appears to again become
kink unstable and erupt (see Fig.\ref{fig:snaps}d)
about $0.4$ hrs after the onset of the first, and as it erupts upward and
pinches off it leaves behind a third newly formed
flux rope that grows quasi-statically (see Fig.\ref{fig:snaps}e).
The process repeats again
for the third eruption (see Figs. \ref{fig:snaps}f and \ref{fig:snaps}g).
All the three eruptions are triggered by the helical kink instability
(see panels (b), (d) and (f) of Fig.\ref{fig:snaps}), with the values
of the decay index $n$ at the apex of the flux rope axis at the onset
of rapid acceleration being $1.0$, $1.05$, and $1.1$ for the first,
second, and third eruptions respectively. These values suggest that all
the three flux ropes are still
stable against the torus instability as they become kink unstable.
Furthermore the configurations shown in panels (c), (e), and (g) of
Fig~\ref{fig:snaps} indicate that all three eruptions are partial eruptions of
the flux rope, with internal reconnections between the two legs of the flux
rope that break the rope in two
\citep[e.g.][]{Tripathi:etal:2009, Gibson:Fan:2006a}.
After the third eruption, a fourth flux rope forms. But in
the course of time this flux rope is found to undergo a sideways
herniation (see Fig. \ref{fig:snaps}h) rather than erupting radially.
Note that each newly formed flux rope remains in a quasi-static
state if the flux emergence is stopped before sufficient twist is transported
in to trigger another kink instability, emphasizing the importance of
sustained flux emergence. 

\begin{figure}[h!]
\includegraphics[width=0.45\textwidth]{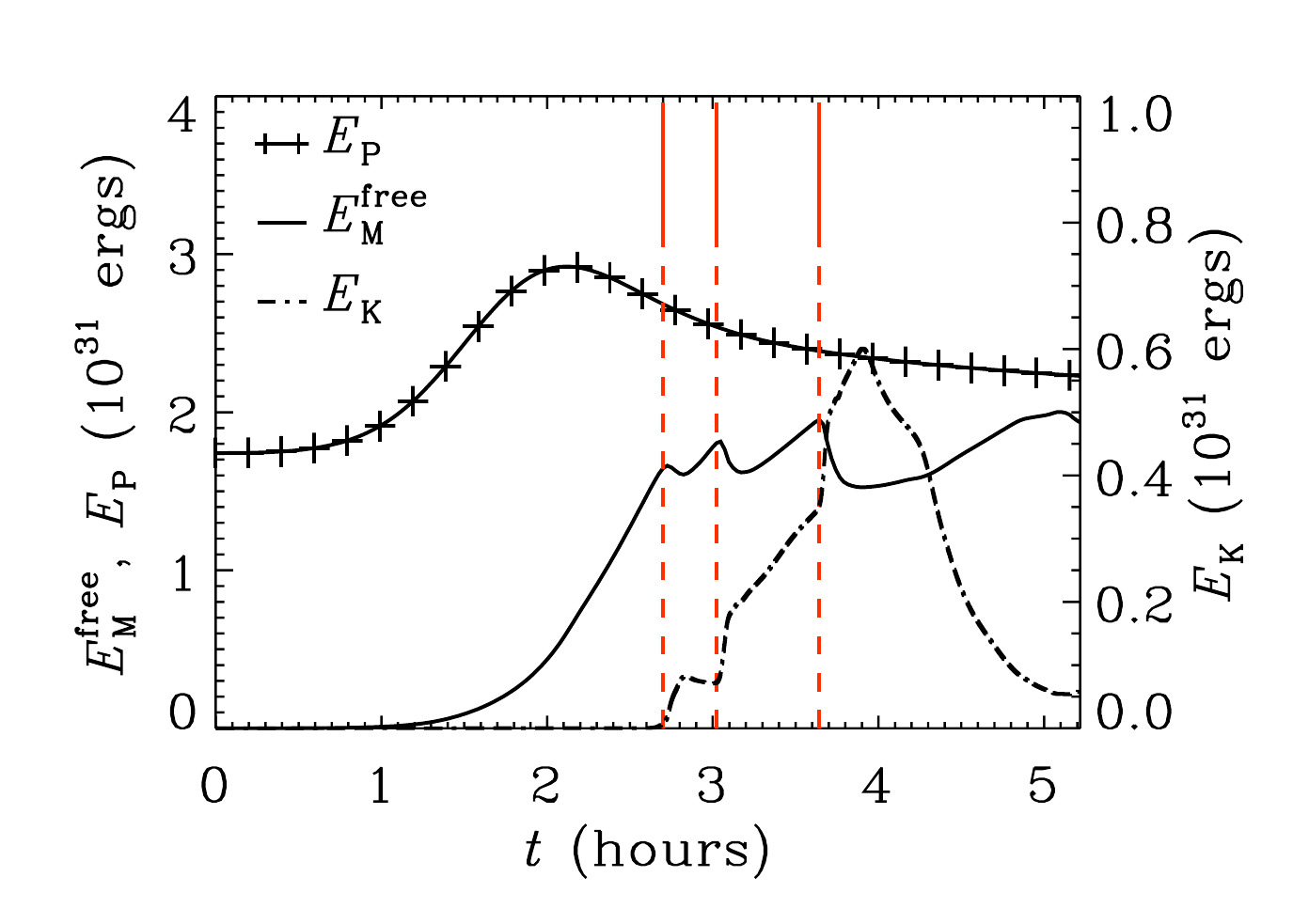}
\caption{\label{fig:energy} Magnetic free energy, $E_{\rm M}^{\rm free}$, $E_{\rm P}$ and kinetic energy, $E_{\rm K}$ as a function of time. The red dashed lines indicate the times for the three CME events.}
\end{figure}

Fig.\ref{fig:energy} shows the temporal evolution of the free magnetic energy, 
$E_{\rm M}^{\rm free}$, defined as the difference between the 
magnetic energy and the energy $E_{\rm P}$ of the corresponding potential
field having the same radial magnetic field distribution on the lower
boundary. $E_{\rm M}^{\rm free}$ measures the maximum available energy for
driving the eruptions.  Also shown in Fig.\ref{fig:energy}
is the evolution of the total kinetic energy $E_{\rm K}$.
The three vertical red dashed lines mark the times for the onset of the three eruptions
characterized by the onset of a sharp increase in $E_{\rm K}$ and
a drop in $E_{\rm M}^{\rm free}$.
Also note for the plot of $E_{\rm K}$ that
the second CME occurs while the first CME blob is still in the domain.
Similarly the third CME occurs when the second and the first are still 
in the domain. Thus the total $E_{\rm K}$ reflects the accumulated kinetic
energy for all three eruptions until about $t=3.9$ hours, when the front of
the leading ejecta begins to exit the domain and $E_{\rm K}$ starts to
decrease.
We find that both the drop in $E_{\rm M}^{\rm free}$ as well as the sharp
increase in $E_{\rm K}$ become progressively greater with each
successive eruption, indicating that each CME is more 
energetic than its predecessor.
A natural consequence of such progressively more energetic successive
eruptions is the cannibalism in homologous CMEs.

\begin{figure*}[h!]
\hspace*{-0.8cm}
\begin{overpic}[width=0.30\textwidth,clip]{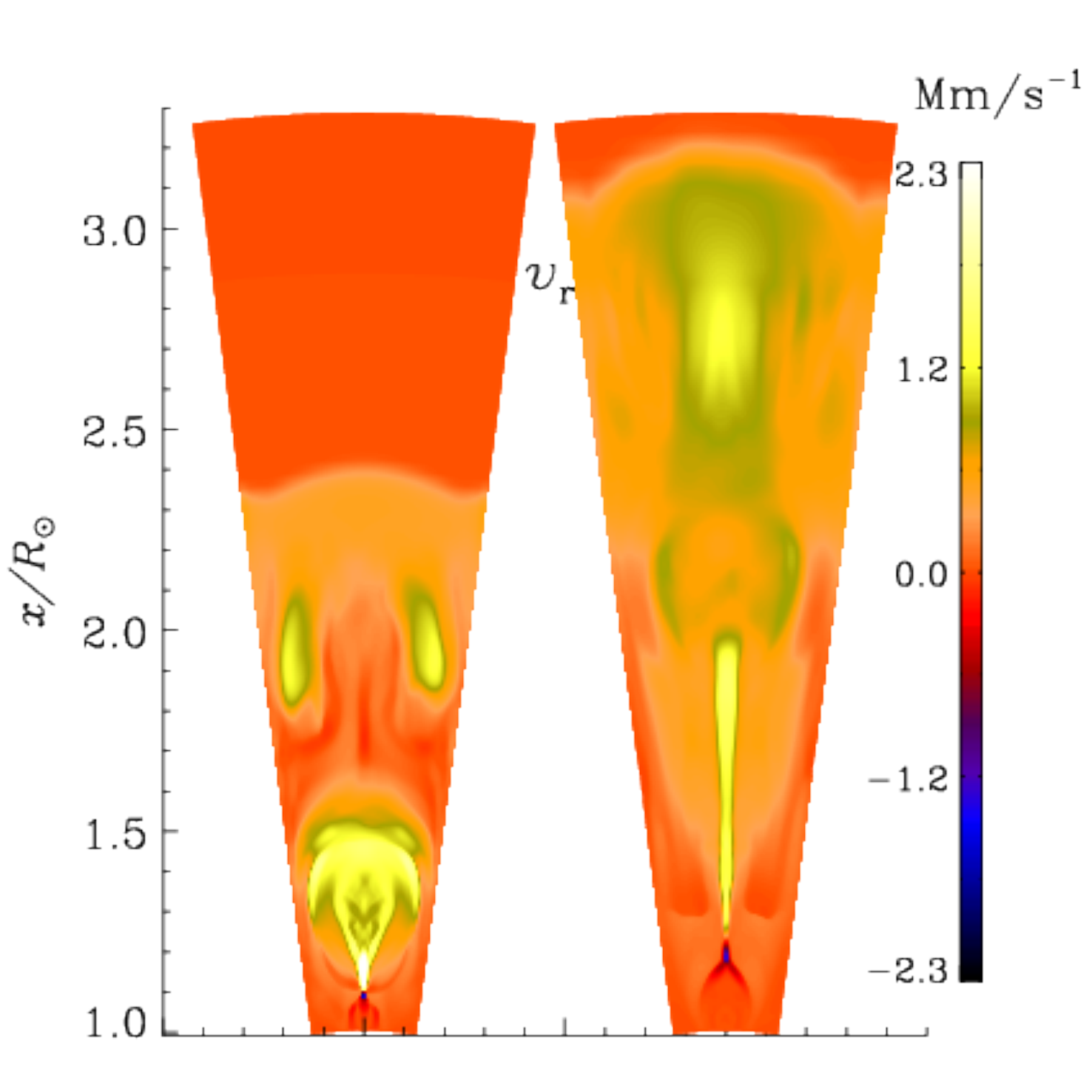}
\put(25,80){(a)}
\end{overpic}
\begin{overpic}[width=0.35\textwidth,clip]{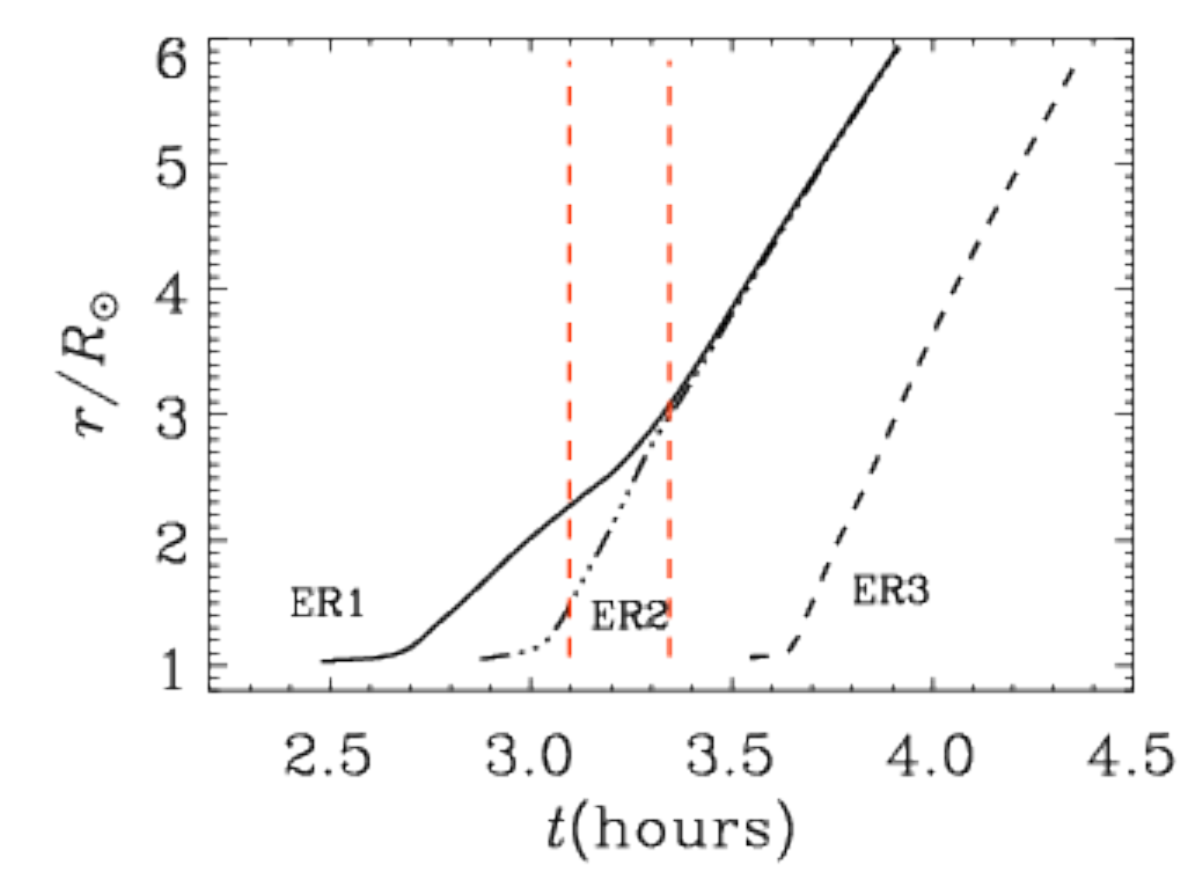}
\put(25,60){(b)}
\end{overpic}
\begin{overpic}[width=0.35\textwidth,clip]{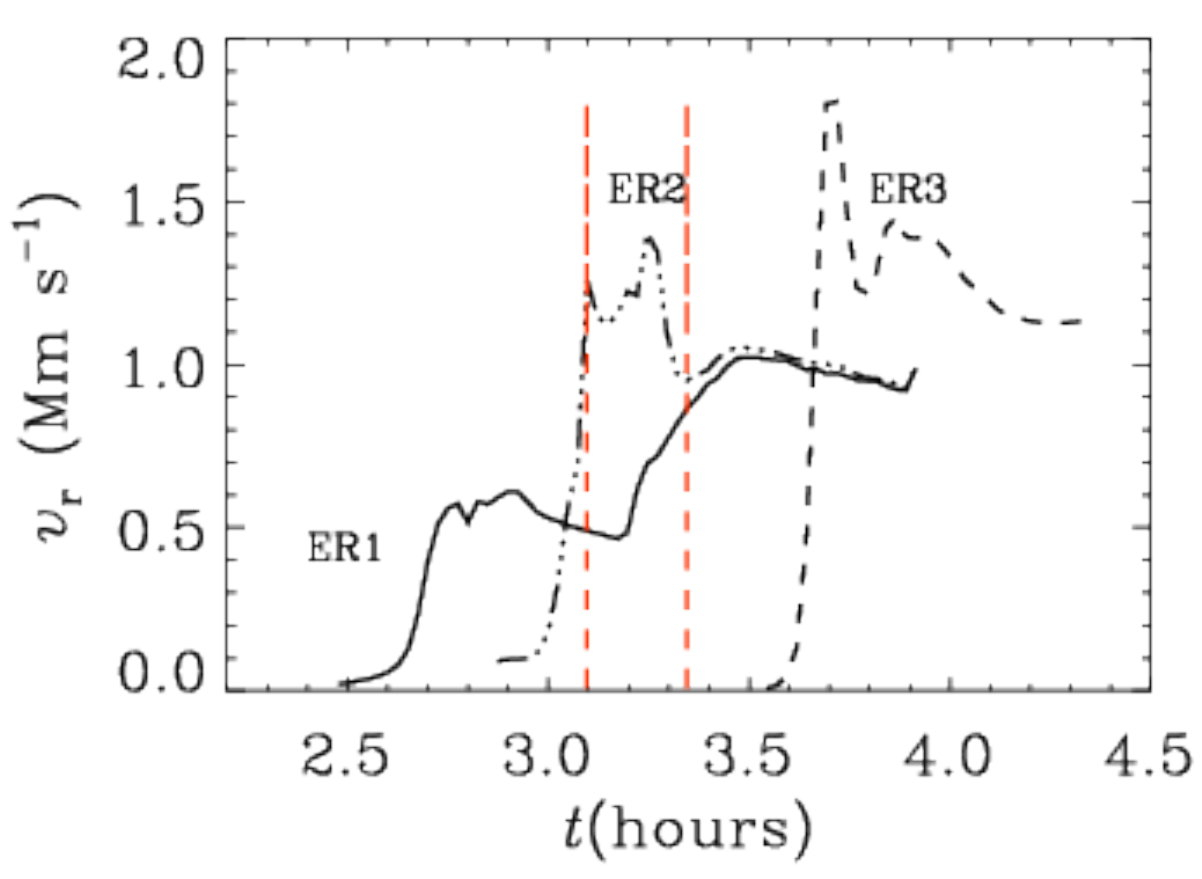}
\put(25,60){(c)}
\end{overpic}
\caption{\label{fig:cannibal} (a) Snapshots of the radial velocity $v_r$ in the central meridional plane across the flux rope at the onset of the second CME (left) and at a time when the second blob catches up with the first (right). A movie showing the evolution of $v_r$ in this plane is available in the online paper . (b) Height vs time of three Lagrangian points (ER1, ER2, ER3) each inside one of the three erupting ropes. (c) Velocity vs time for the same Lagrangian points tracked in (b). The red dashed lines in (b) and (c) indicate the exact time of the snapshots in (a).}
\end{figure*}

Fig.\ref{fig:cannibal}a show two snapshots of the radial velocity in the
central cross-section. The earlier snapshot shows the second CME erupting while the front of the first ejecta has already reached $~2.4 R_{\odot}$. The latter
snapshot shows the time just after the second CME blob has ``gobbled up'' the
first blob, forming a single fast ejecta moving outward at a speed of
$\sim 1$ Mm/s. A movie showing this evolution of $v_r$ in the cross-section
is available in the online version.  More quantitatively, Figs. \ref{fig:cannibal}b and \ref{fig:cannibal}c show respectively the height vs. time and $v_r$ vs.
time for three Lagrangian points (ER1, ER2 and ER3), each tracked starting
from the apex of each of the three flux ropes at the onset of its eruption. 
Fig. \ref{fig:cannibal}b shows that point ER2 catches up with ER1 at $t=3.35$
hours. The point ER3 is fastest of the three but does not catch up with
ER1 and ER2 inside the domain. If the radial extent of our domain had been
much longer, we might have witnessed another event of cannibalism.
It is clear from Fig.\ref{fig:cannibal}c that ER1, ER2 and ER3 exhibit
increasingly stronger acceleration, reaching peak velocities of
$650$ km s$^{-1}$, $1400$ km s$^{-1}$ and  $1800$ km s$^{-1}$ respectively.
All three CMEs in our simulation can be classified as fast CMEs. The greater
acceleration of the following CME compared to its precursor is because the
following flux rope is erupting into the field that has been opened up by the
leading eruption and therefore has less downward magnetic tension to overcome.
However at the instant of collision between the first and the second CME the
merged ejecta attains a speed that is greater than the first CME but slower
than the second in order to conserve momentum. The merged CME exits the
domain traveling at a speed of $~950$ km s$^{-1}$.

\begin{figure*}[h!]
\begin{overpic}[width=0.225\textwidth,clip]{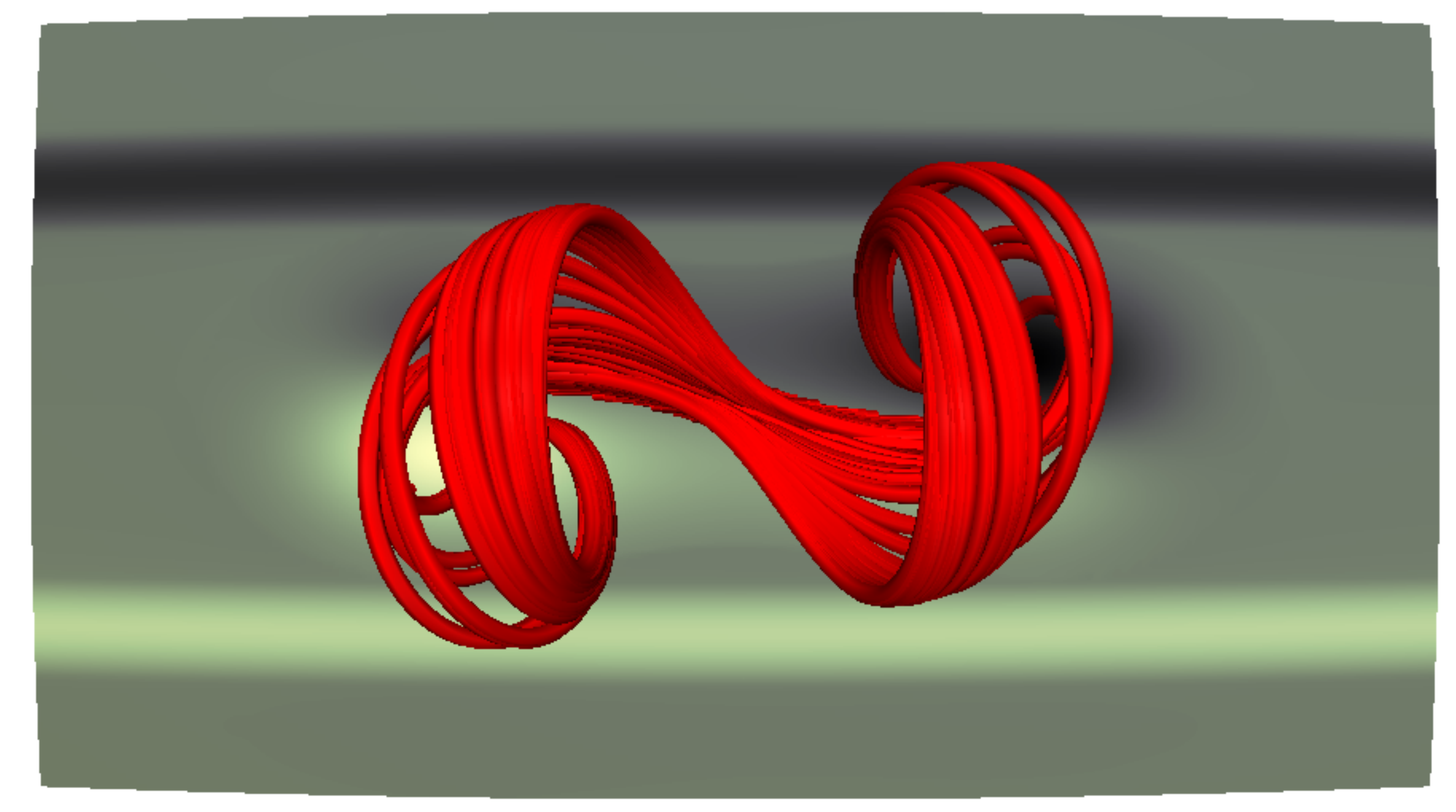}
\put(5,90){(a) $t=2.67$}
\end{overpic}
\begin{overpic}[width=0.225\textwidth,clip]{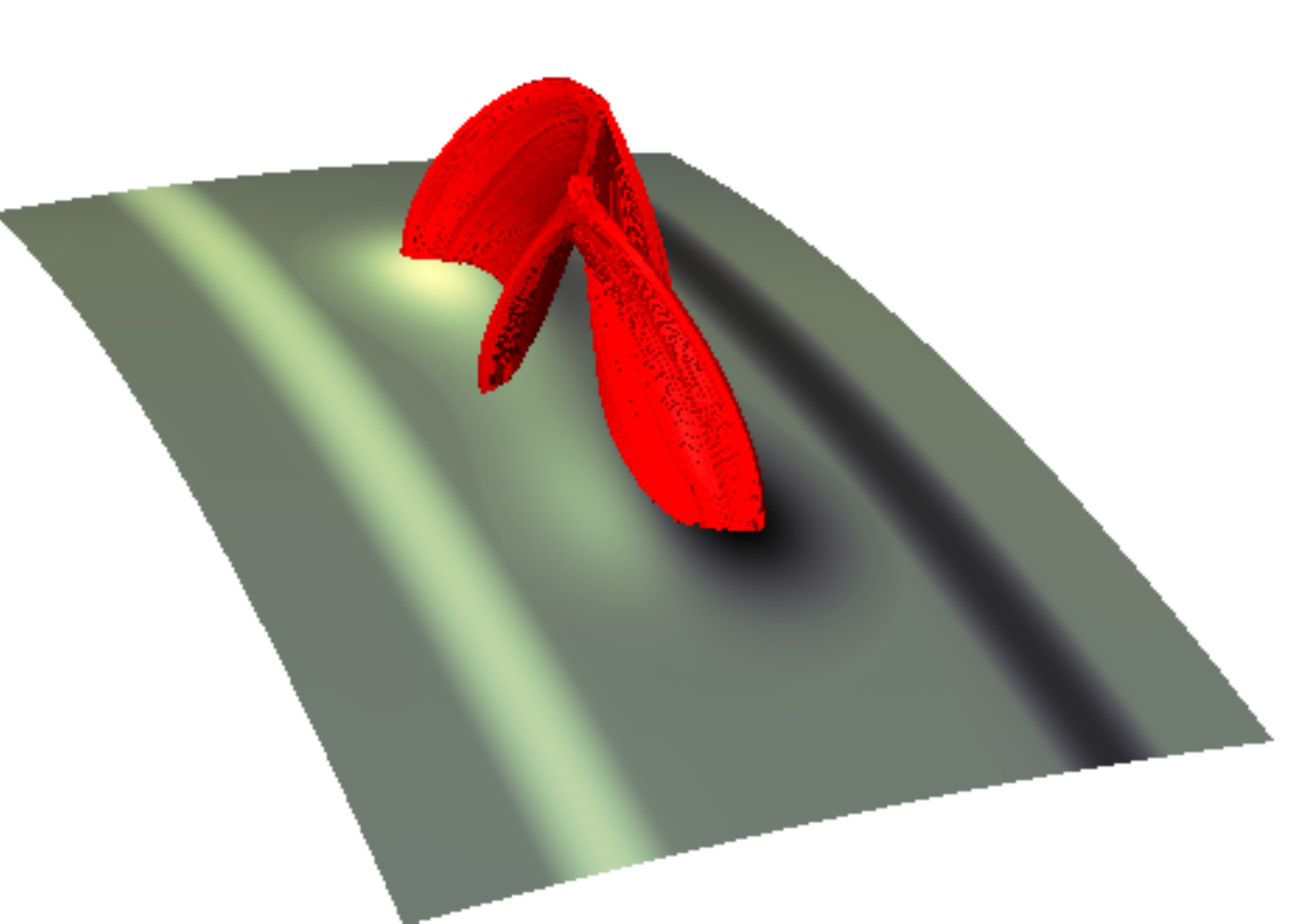}
\put(5,90){(b) $t=2.72$}
\end{overpic}
\begin{overpic}[width=0.225\textwidth,clip]{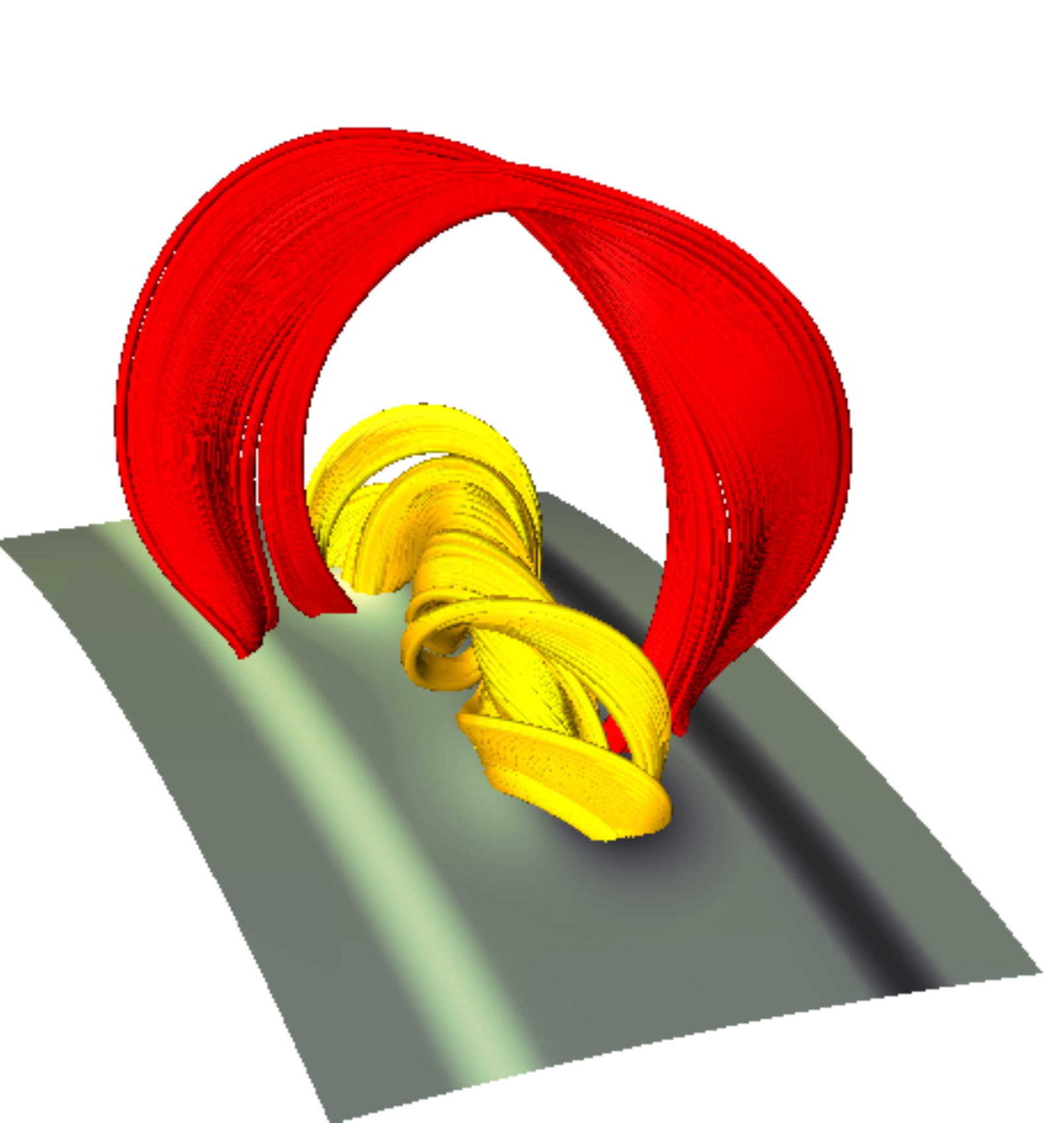}
\put(5,90){(c) $t=2.87$}
\end{overpic}
\begin{overpic}[width=0.225\textwidth,clip]{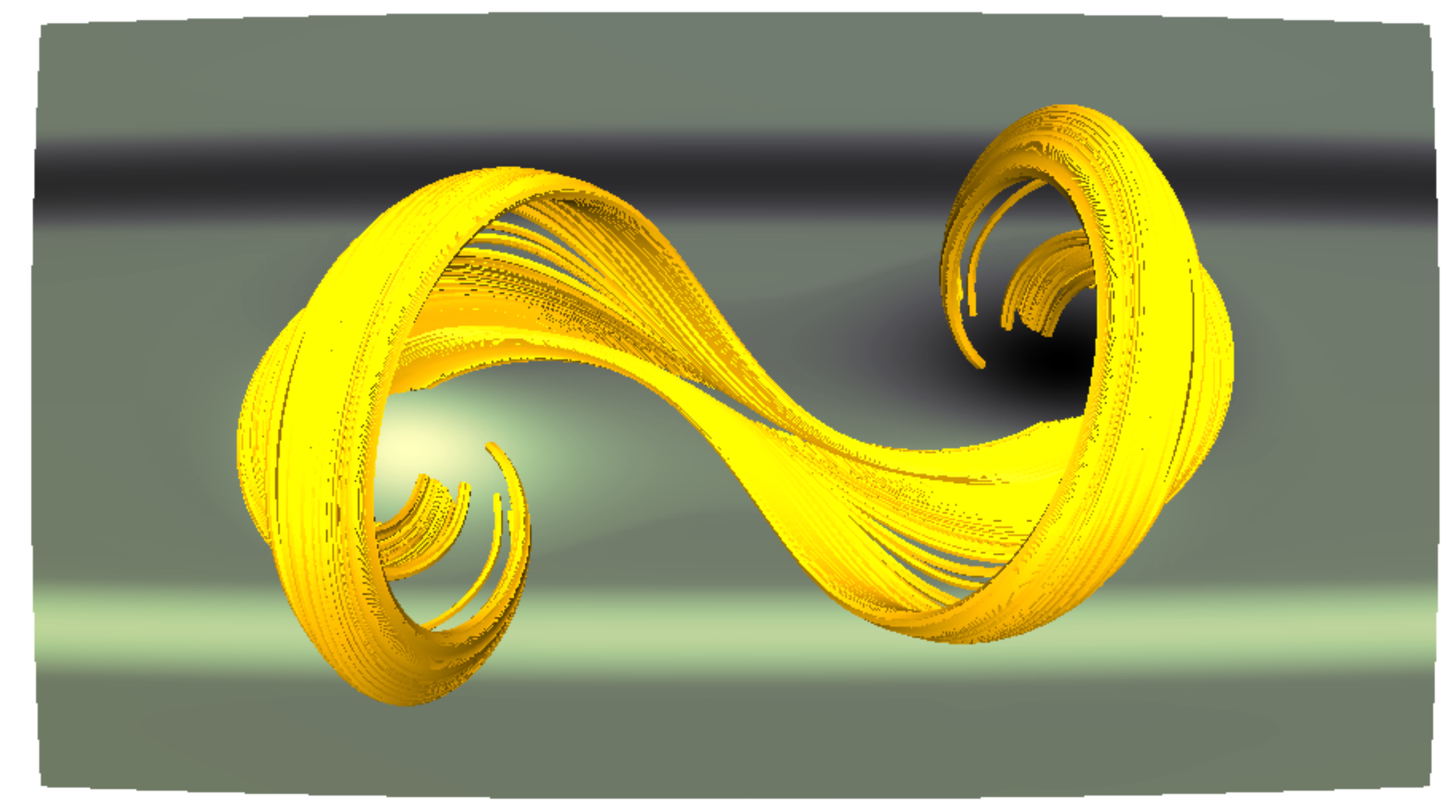}
\put(5,90){(d) $t=3.02$}
\end{overpic}
\caption{\label{fig:cs_sig} The transition in the morphology of the 
heated field lines from a sigmoid shape to cusped post flare 
loops and back to sigmoid. The red fieldlines belong 
to the first erupting rope where as the yellow fieldlines belong to the second 
erupting rope. The sigmoid fieldlines in panels (a) (in red),
and (c) and (d) (in yellow), are traced from the current layer isosurface
with a value of $J/B \sim 1/8\delta x$ whereas the cusped fieldlines in
panels (b) and (c) (in red), are traced from the current layer isosurface
with a value $J/B \sim 1/3.5\delta x$. Time is indicated in hours.}
\end{figure*}

The repeated reformation of the coronal flux rope after each eruption may be identified with the repeated reformations of the X-ray sigmoids in the active region.
Fig.\ref{fig:cs_sig} shows snapshots of the morphology of the
most heated field lines between the first and the second eruptions to
illustrate this point.
These most heated field lines are selected by tracing field lines from the
points in the thin current layers with $J/B$ above a selected high
value, which is $\sim 1/8 \delta x$ for sigmoid fieldlines in panels (a), (c) and (d), 
and $\sim 1/3.5 \delta x$ for cusped fieldlines of panels (b) and (c) in Fig.\ref{fig:cs_sig},
where $J$, $B$, and $\delta x$ denote the current density,
the field strength, and the minimum grid spacing respectively.
Panel (a) shows the most heated field lines (the red field lines) at the
onset of the kink instability of the first flux rope, and they show an
inverse $\mathcal{S}$ morphology as expected for a left-hand-twisted
flux rope \citep[e.g.][]{Low:Berger:2003,Fan:Gibson:2004}.
As the eruption of the first flux rope progresses, the current sheet
intensifies and rapid reconnections take place.
The heated field lines traced from the most intense part
(with $J/B > 1/3.5 \delta x$) of the current sheet form the cusped post
flare loops as shown in Fig.\ref{fig:cs_sig}b.
As the current sheet stretches outward, the foot points of the
post flare loop widens (red field lines in Fig.\ref{fig:cs_sig}c).
A new sigmoid traced by a family of yellow colored field lines has
appeared below the cusped post flare loop as shown in Fig.\ref{fig:cs_sig}c. 
The formation of the second flux rope is aided by the magnetic reconnections in the 
partial eruption of the first flux rope, as flux emergence continues across the lower boundary.
This is the familiar ``sigmoid-under-cusp'' configuration so often seen in X-ray observations of the 
homologous-CME source regions \citep[e.g.][]{Gibson:etal:2002}. Finally the current sheet through which 
we trace the cusped post flare loops dissipates away leaving behind a second well formed sigmoid 
(Fig.\ref{fig:cs_sig}d). This pattern of sigmoid-cusp-sigmoid repeats itself during each CME observed in our simulation.

\section{Discussion}
We have demonstrated with an MHD simulation that cannibalistic, homologous CMEs may result from 
repeated formations and partial eruptions of kink unstable flux ropes during the course of a 
sustained emergence of highly twisted magnetic fields in an active region. The first of a sequence 
of homologous CMEs must perform work to open up a confining field that is closed ahead of it.  The 
subsequent CMEs tend to be cannibalistic because each is erupting into the already opened field 
stretched out by its preceding CME, the former thus being able to accelerate to a higher speed to 
catch up and merge into the latter. The speeds of the CMEs in our simulation are  $\sim 1000$ km s$^{-1}$. 
A merged CME produced by such a cannibalistic process is naturally fast and massive, and, therefore, 
geo-effective.  The essential point is about the MHD conversion of total free magnetic energy 
available in the highly-twisted, complex magnetic fields that continually emerge and create multiple 
CMEs sufficiently close in time for a few of them to eventually merge cannibalistically. 
A significant part of that free energy is converted into the terminal kinetic energy by this process. 
If these CMEs were to be individually created in separate events, each involving opening up and 
reclosing of an initially closed field, a significant amount of the free magnetic energy must go 
into opening up the confining field.  
This non-trivial amount of energy \citep{Aly:1987} does not become a 
part of the terminal CME kinetic energy because it is to be liberated in a post-CME flare associated 
with the re-closing of the open field.  We will study the detailed magnetic field structure and 
topology of the merged CMEs in a subsequent paper. Our current simulation
has only considered highly twisted emerging flux ropes that develop the kink
instability. 
There may also exist other mechanisms for triggering multiple eruptions, 
e.g. breakout reconnections and the onset of the torus instability of the flux rope \citep{MacTaggart:Hood:2009}.
A detailed parametric study with varying twist of the emerging flux
rope is needed to study how the development of homologous CMEs depends on the onset
of the kink instability.

\acknowledgments
We thank B. C. Low for his comments and revisions which 
have significantly improved the manuscript. 
This work is supported by NASA LWS grant NNX09AJ89G to NCAR.
NCAR is sponsored by the National Science Foundation.
The numerical simulations were carried out on
the Yellowstone supercomputer of NWSC/NCAR under the NCAR Strategic
Capability computing project NHAO0001, and also on the Discover supercomputer
at NASA Center for Climate Simulation under the project GID s0969.

\end{document}